\documentclass[12pt]{article}
\usepackage{subfloat}
\pdfoutput = 1
\usepackage{graphics}
\usepackage{graphicx} 
\begin{document}
\date{Today}
\title{{\bf{\Large  Noncommutative effects of spacetime on holographic superconductors }}}

\author{
{\bf {\normalsize Debabrata Ghorai}$^{a}$
\thanks{debanuphy123@gmail.com, debabrataghorai@bose.res.in}},\,
{\bf {\normalsize Sunandan Gangopadhyay}$^{b,c}
$\thanks{sunandan.gangopadhyay@gmail.com}}\\
$^{a}$ {\normalsize  S.N. Bose National Centre for Basic Sciences,}\\{\normalsize JD Block, 
Sector III, Salt Lake, Kolkata 700098, India}\\[0.2cm]
$^{b}$ {\normalsize Department of Physics, West Bengal State University, Barasat, India}\\
$^{c}${\normalsize Visiting Associate in Inter University Centre for Astronomy \& Astrophysics,}\\
{\normalsize Pune, India}\\[0.1cm]
}
\date{}

\maketitle

\begin{abstract}
{\noindent The Sturm-Liouville eigenvalue method is employed to analytically investigate the properties of holographic superconductors in higher dimensions in the framework of Born-Infeld electrodynamics incorporating the effects of noncommutative spacetime.
In the background of pure Einstein gravity in noncommutative spacetime, we obtain the relation between the critical temperature and the charge density. We also obtain the value of the condensation operator and the critical exponent. Our findings suggest that higher the value of noncommutative parameter and Born-Infeld parameter make the condensate harder to form. We also observe that the critical temperature depends on the mass of the black hole and higher value of black hole mass is favourable for the formation of the condensate.  }
\end{abstract}
\vskip 1cm

\section{Introduction}
\noindent Holographic superconductors have been studied extensively in recent times. Their importance lies in the fact that they mimic some properties of high $T_c$ superconductors. The interest rose after the demonstration in \cite{hs2} that an Abelian Higgs model in AdS spacetime leads to a spontaneous symmetry breaking and thus giving rise to a scalar hair near the horizon of the black hole. The important ingredient which goes in the construction of such holographic superconductor models is the correspondence between gravity and gauge theory, namely,the AdS/CFT correspondence \cite{adscft1}. 

\noindent Spacetime noncommutativity has been another prominent area of research in recent years. The idea of noncommutative (NC) spacetime, first formally introduced by Snyder \cite{ncex1} back in 1947 was not considered seriously by other scientist till recently when such a structure emerged naturally from investigations carried out in string theory \cite{ncex2}. It was in this paper that NC field theory was resurrected and rules were given to move from ordinary quantum field theory (QFT) to NC QFT. In more recent times, a noncommutative inspired Schwarzschild metric was obtained in \cite{nico1,nico2}. Here the effect of noncommutativity was introduced through a smeared matter source which was then used to solved Einstein's equation of general relativity. An important aspect of this black hole solution was the removal of black hole singularity. The thermodynamics of this black hole solution was investigated in details in \cite{rb1}.

\noindent In this paper, we want to investigate the role of NC geometry on the AdS/CFT duality, in particular to study its effect on holographic superconductor models in higher dimensions. Such a study had been carried out earlier in \cite{subir} in $4-$dimensions. Here we generalize this analysis to arbitrary dimensions by considering the $d-$dimensional generalization of the NC Schwarzschild black hole. We also present expressions for the critical temperature which is more accurate than the one given in \cite{subir} as will be clear in the subsequent discussion. Further we consider Born-Infeld (BI) electrodynamics thereby including the effect of non-linearity in the analysis. There are quite a few reasons which makes it worthwhile to study the effect of BI electrodynamics. First of all, it is the only non-linear theory that remains invariant under electromagnetic duality. Another intriguing feature is that it has a nice weak field limit \cite{ncex8} - \cite{ncex7}. It also finds remarkable application in string theory \cite{ncex3,ncex4}. We would like to mention that the technique that we have adopted in this paper to obtain the relation between the critical temperature and the charge density is the Sturm-Liouville (SL) eigenvalue approach.

\noindent This paper is organized as follows. In section 2, we show the basic holographic set up in noncommutative spcetime in the background of electrically charged black hole in arbitrary dimension. In section 3, taking into account the effect of the Born-Infeld electrodynamics, we have derived the relation between critical temperature and charge density using the Sturm-Liouville eigenvalue problem. In section 4, we analytically obtain an expression for the condensation operator in $d$-dimension near the critical temperature. We conclude finally in section 5.


\section{Set up in noncommutative spacetime}
We by considering a noncommutative charged Schwarzschild-$AdS_d$ black hole whose metric is given by \cite{nico2} 
\begin{eqnarray}
ds^2=-f(r)dt^2+\frac{1}{f(r)}dr^2+ r^2 h_{ij} dx^{i} dx^{j} \nonumber \\
f(r)= K+ \frac{r^2}{L^2} -\frac{2MG_d}{r^{d-3}\Gamma(\frac{d-1}{2})} \gamma\left(\frac{d-1}{2}, \frac{r^2}{4\theta}\right)
\label{m1}
\end{eqnarray}
where $ h_{ij} dx^{i} dx^{j}$ denotes the line element of a $(d-2) $-dimensional hypersurface with zero curvature and 
\begin{eqnarray}
\gamma(s,x)= \int^{x}_{0} t^{s-1} e^{-t} dt
\label{gammanc}
\end{eqnarray}
is the lower incomplete Gamma function and $K$ represents the curvature. As $\theta\rightarrow 0$, the noncommutative metric $f(r)$ gives back the commutative Schwarzschild metric in $d-$dimensions. We also work in the probe limit which basically implies that the backreaction on the spacetime metric $f(r)$ is not taken into account. 

\noindent The Hawking temperature of this black hole, which is interpreted as the temperature of the conformal field theory on the boundary, is given by
\begin{eqnarray}
T_{H} = \frac{f^{\prime}(r_{+})}{4\pi}
\label{gzx1}
\end{eqnarray} 
where $ r_{+} $ is the radius of the horizon of the black hole. \\
Since the construction of the holographic $s-$wave superconductor requires a planar symmetry, we set $K=0$ which implies that  
\begin{eqnarray}
f(r) = \frac{r^2}{L^2} -\frac{2MG_d}{r^{d-3}\Gamma(\frac{d-1}{2})} \gamma\left(\frac{d-1}{2},\frac{r^2}{4\theta}\right).
\label{metricnc}
\end{eqnarray} 
Using the fact that metric vanishes at the event horizon, we get the radius of the event horizon $r_+$ in $d-$dimensions to be
\begin{eqnarray}
r^{d-1}_{+} = \frac{2MG_d L^2}{\Gamma(\frac{d-1}{2})} \gamma\left(\frac{d-1}{2},\frac{r^{2}_{+}}{4\theta}\right).
\label{r1nc}
\end{eqnarray}
For convenience we shall set $L=1$ in the rest of the analysis. The above relation enables us to write the metric (\ref{metricnc}) as 
\begin{eqnarray}
f(r)=r^2 - \frac{r^{d-1}_{+}}{r^{d-3}}.\frac{\gamma(\frac{d-1}{2},\frac{r^2}{4\theta})}{\gamma(\frac{d-1}{2},\frac{r^{2}_{+}}{4\theta})}.
\label{nchscf1}
\end{eqnarray}
From eq.(s) (\ref{gzx1}) and (\ref{nchscf1}), we obtain the expression for the Hawking temperature of the black hole 
\begin{eqnarray}
T_{H} &=& \frac{1}{4\pi}\left[ (d-1)r_{+} - r^2_{+} \frac{\gamma^{\prime}(\frac{d-1}{2}, \frac{r^2_{+}}{4\theta})}{\gamma(\frac{d-1}{2}, \frac{r^2_{+}}{4\theta})}. \right]
\end{eqnarray}
Computing the derivative of the incomplete Gamma function (\ref{gammanc}), we get
\begin{eqnarray}
T_H = \frac{r_+}{4\pi} \left[(d-1) - \frac{4MG_d}{\Gamma(\frac{d-1}{2})}.\frac{e^{-\frac{r^2_+}{4\theta}}}{(4\theta)^{\frac{d-1}{2}}}\right]
\label{de9} 
\end{eqnarray}

The matter Lagrangian density is due to the presence of a gauge field and complex scalar field. This reads
\begin{eqnarray}
\mathcal{L}_{matter}= \frac{1}{b}\left(1-\sqrt{1+ \frac{b}{2}F^{\mu \nu} F_{\mu \nu}}\right) - (D_{\mu}\psi)^{*} D^{\mu}\psi-m^2 \psi^{*}\psi
\label{ac2}
\end{eqnarray}
where $F_{\mu \nu}=\partial_{\mu}A_{\nu}-\partial_{\nu}A_{\mu}$; ($\mu,\nu=0,1,2,3,4$) is the field strength tensor, 
$D_{\mu}\psi=\partial_{\mu}\psi-iqA_{\mu}\psi$ is the covariant derivative,  $A_{\mu}$ and $ \psi $ represent the gauge field and the scalar field respectively.\\
\noindent Considering that the black hole possesses only electric charge, we make the ansatz \cite{hs6} $A_{\mu} = (\phi(r),0,0,0)$ and $ \psi=\psi(r)$. We can also choose magnetic field but now we only have presented electric field. Using this ansatz, the equations of motion for the scalar fields and electric potential read \cite{dg1}
\begin{eqnarray}
\psi^{\prime \prime}(r) + \left(\frac{d-2}{r} + \frac{f^{\prime}(r)}{f(r)}\right)\psi^{\prime}(r) + \left(\frac{q^2 \phi^{2}(r)}{f(r)^2}- \frac{m^{2}}{f(r)}\right)\psi(r) = 0
\label{e01}
\end{eqnarray}
\begin{eqnarray}
\phi^{\prime \prime}(r) + \frac{d-2}{r} \phi^{\prime}(r) - \frac{d-2}{r} b \phi^{\prime}(r)^{3} - \frac{2 q^2 \phi(r) \psi^{2}(r)}{f(r)}(1 - b \phi^{\prime}(r)^{2})^\frac{3}{2} = 0
\label{e1}
\end{eqnarray}
where prime denotes derivative with respect to $r$. 
The rescalings $\psi\rightarrow \psi/q$, $\phi\rightarrow \phi/q$ and $\kappa^2 \rightarrow q^2 \kappa^2 $ \cite{betti} allows one to set $q=1$.

\noindent To solve the above non-linear coupled differential equations (\ref{e01})-(\ref{e1}), we must fix the boundary condition for $\phi(r)$ and $\psi(r)$ which are physically acceptable. For regularization, one requires $\phi(r_+)=0$ and $\psi(r_{+})$ to be finite at the horizon. 

\noindent Near the boundary of the bulk, the asymptotic behaviour of $\psi$ and $\phi$ are not affected by noncommutativity. This is because near the boundary, $r$ is large and therefore $e^{\frac{-r^2}{4\theta}}\ll 1$ since $\theta$ is small. The asymptotic behaviour of the fields can be written as \cite{hs8}
\begin{eqnarray}
\label{bound1}
\phi(r)&=&\mu-\frac{\rho}{r^{d-3}}\\
\psi(r)&=&\frac{\psi_{-}}{r^{\Delta_{-}}}+\frac{\psi_{+}}{r^{\Delta_{+}}}
\label{bound2}
\end{eqnarray}
where
\begin{eqnarray}
\label{bound1a}
\Delta_{\pm}&=&\frac{(d-1)\pm\sqrt{(d-1)^2+4m^2}}{2}~.
\end{eqnarray}
The gauge/gravity duality allows one to interpret $\rho$ and $\mu$ as the charge density and chemical potential of the boundary field theory. For the choice $m^2 =-3$ with the Breitenlohner-Freedman bound \cite{ncex9}, we have $\Delta_{+} =3 ~\Delta_{-} =1$ for $d=5$.
This allows one to choose $\psi_{+}$ or $\psi_{-}$. In this paper we shall choose $\psi_{-}=0$. This basically means that $\psi_{+}$ is dual to the expectation value of the condensation operator $J$ in the absence of the source $\psi_{-}$.

\noindent Using $z=\frac{r_{+}}{r}$,  the field equations (\ref{e01})-(\ref{e1}) takes the form
\begin{eqnarray}
\psi^{\prime \prime}(z) + \left(\frac{f^{\prime}(z)}{f(z)} - \frac{d-4}{z}\right)\psi^{\prime}(z) + \frac{r^2_{+}}{z^4} \left(\frac{\phi^{2}(z)}{f(z)^2}- \frac{m^{2}}{f(z)}\right)\psi(z) =0 
\label{e1bb}
\end{eqnarray}
\begin{eqnarray}
\phi^{\prime \prime}(z) -\frac{d-4}{z} \phi^{\prime}(z) + \frac{d-2}{r^2_{+}} b \phi^{\prime}(z)^{3} z^3 - \frac{2r^2_{+} \phi(z) \psi^{2}(z)}{f(z) z^4}\left(1 -\frac{b z^4 }{r^2_{+}} \phi^{\prime}(z)^{2}\right)^\frac{3}{2} = 0
\label{e1aa}
\end{eqnarray}
where prime now denotes derivative with respect to $z$. These equations are to be solved in the interval $(0, 1)$, where $z=1$ is the horizon and $z=0$ is the boundary.
The boundary condition $\phi(r_+)=0$ now becomes $\phi(z=1)=0$.



\section{Critical temperature ($T_{c}$) and charge density ($\rho$)}
\noindent To begin the analysis, we first note that for $T \geq T_c $, the matter field must vanish. We want to study the behaviour of $\psi(r)$ just below the critical temperature $(T_c)$. For that we first need to solve the $\phi(r)$ at $T=T_c$, where $\psi(r)$ vanishes. Hence eq.(\ref{e1aa}) reduces to 
\begin{eqnarray}
\label{metric1}
\phi^{\prime \prime}(z) -\frac{d-4}{z}\phi^{\prime}(z) + \frac{(d-2)bz^3}{r^2_{+(c)}} \phi^{\prime}(z)^3=0.
\end{eqnarray}
We now employ the perturbative technique developed in \cite{sgdc1} to solve the above equation. When $b=0$, the solution of the above equation (compatible the boundary condition of $\phi(z)$) reads
\begin{eqnarray}
\phi(z)|_{b=0}= \lambda r_{+(c)} (1-z^{d-3})
\label{de1}
\end{eqnarray}
where
\begin{eqnarray}
\lambda=\frac{\rho}{r_{+(c)}^{d-2}}~.
\label{lam}
\end{eqnarray}
To solve eq.(\ref{metric1}), we put the solution for $\phi(z)$ with $b=0$ (i.e. $\phi(z)|_{b=0}$) in the non-linear term of 
eq.(\ref{metric1}). This leads to
\begin{eqnarray}
\phi^{\prime \prime}(z) -\frac{d-4}{z} \phi^{\prime}(z) - b\lambda^3 r_{+(c)} (d-2)(d-3)^3 z^{3(d-3)} =0.
\label{de2}
\end{eqnarray}    
Using the asymptotic boundary condition (\ref{bound1}), the solution of the above equation 
upto first order in the Born-Infeld parameter $b$ reads 
\begin{eqnarray}
\phi(z) = \lambda r_{+(c)}\left\{ (1-z^{d-3}) - \frac{b(\lambda^2|_{b=0}) (d-3)^3}{2(3d-7)} (1-z^{3d-7})\right\}
\label{de3} 
\end{eqnarray}
where we have used the fact that $b\lambda^2= b(\lambda^2|_{b=0}) + \mathcal{O}(b^2)$ \cite{sgdc1}, $\lambda^2|_{b=0}$ being the value of $\lambda^2$ for $b=0$.

\noindent Under change of coordinate $z=\frac{r_+}{r}$ and $T = T_c$, the metric (\ref{nchscf1}) reads 
\begin{eqnarray}
f(z)= \frac{r^2_{+(c)}}{z^2} g_{0}(z)
\label{de7}
\end{eqnarray}
where 
\begin{eqnarray}
g_{0}(z) & = & 1- z^{d-1}.\frac{\gamma(\frac{d-1}{2},\frac{r^2_{+(c)}}{4\theta z^2})}{\gamma(\frac{d-1}{2},\frac{r^{2}_{+(c)}}{4\theta})} \nonumber \\
&\approx & 1- z^{d-1} \Xi
\label{metr33}
\end{eqnarray} 
where $\Xi =1+\frac{r^{d-3}_{+(c)}}{\Gamma (\frac{d-1}{2})}.\frac{e^{-\frac{r^2_{+(c)}}{4\theta}}}{(4\theta)^{\frac{d-3}{2}}} + \frac{d-3}{2}.\frac{r^{d-5}_{+(c)}}{\Gamma (\frac{d-1}{2})}.\frac{e^{-\frac{r^2_{+(c)}}{4\theta}}}{(4\theta)^{\frac{d-5}{2}}} + \frac{(d-3)(d-5)}{2^2}.\frac{r^{d-7}_{+(c)}}{\Gamma (\frac{d-1}{2})}.\frac{e^{-\frac{r^2_{+(c)}}{4\theta}}}{(4\theta)^{\frac{d-7}{2}}} +...... $\\

\noindent Note that for $d=5$, the terms proportional to $(d-5)$ vanish. Also we have neglected the terms of the form $e^{-\frac{r^2_{+(c)}}{4\theta z^2}}$ as they are small compared to $e^{-\frac{r^2_{+(c)}}{4\theta}}$.

\noindent Near the critical temperature $T\rightarrow T_c$, eq.(\ref{e1bb}) for the field $\psi$ approaches the limit
\begin{eqnarray}
\psi''(z)+ \left(\frac{g'_{0}(z)}{g_{0}(z)}-\frac{d-2}{z}\right)\psi'(z)+ \left( \frac{\phi^{2} (z)}{g^{2}_{0} (z) r^{2}_{+(c)}} - \frac{m^2}{g_{0}(z) z^2}\right)\psi(z)=0
\label{e001}
\end{eqnarray}
where $\phi(z)$ now corresponds to the solution in eq.(\ref{de3}). 

Define \cite{siop}
\begin{eqnarray}
\psi(z)=\frac{\langle J\rangle}{r^{\Delta_{+}}_{+}} z^{\Delta_{+}} F(z)
\label{sol1}
\end{eqnarray}
near the boundary, where $F(0)=1$ and $J$ is the condensation operator and substituting this form of $\psi(z)$ in eq.(\ref{e001}), we obtain
\begin{eqnarray}
F''(z) &+& \left\{\frac{2\Delta_{+} -d+2}{z} -\frac{(d-1).\Xi. z^{d-2}}{1-\Xi. z^{d-1}} \right\}F'(z) \nonumber \\ &+&\left\{ \frac{\Delta_{+}(\Delta_{+} -1)}{z^2} -\left(\frac{(d-1).\Xi. z^{d-2}}{1-\Xi. z^{d-1}}+\frac{d-2}{z}\right)\frac{\Delta_{+}}{z}-\frac{m^2}{(1-\Xi. z^{d-1}) z^2} \right\}F(z) \nonumber \\
&+& \frac{\lambda^2}{(1-\Xi. z^{d-1})^2}\left\{ (1-z^{d-3})^2 -\frac{b(\lambda^2|_{b=0})(d-3)^3}{3d-7}(1-z^{d-3})(1-z^{3d-7})\right\}F(z)=0 \nonumber \\
\label{eq5b}
\end{eqnarray}
to be solved subject to the boundary condition $F' (0)=0$. 
 
\noindent The above equation can be recast in the Sturm-Liouville form 
\begin{eqnarray}
\frac{d}{dz}\left\{p(z)F'(z)\right\}+q(z)F(z)+\lambda^2 r(z)F(z)=0
\label{sturm}
\end{eqnarray}
with 
\begin{eqnarray}
p(z)&=&z^{2\Delta_{+} -d+2}(1-\Xi. z^{d-1})\nonumber\\
q(z)&=&z^{2\Delta_{+} -d+2}(1-\Xi. z^{d-1})\left\{ \frac{\Delta_{+}(\Delta_{+} -1)}{z^2} -\left(\frac{(d-1).\Xi. z^{d-2}}{1-\Xi. z^{d-1}}+\frac{d-2}{z}\right)\frac{\Delta_{+}}{z}-\frac{m^2}{(1-\Xi. z^{d-1}) z^2} \right\} \nonumber\\
r(z)&=&\frac{z^{2\Delta_{+} -d+2}}{1-\Xi. z^{d-1}} \left\{ (1-z^{d-3})^2 -\frac{b(\lambda^2|_{b=0}) (d-3)^3}{3d-7}(1-z^{d-3})(1-z^{3d-7})\right\} .
\label{i1}
\end{eqnarray}
To estimate the eigenvalue $\lambda^2$, we write down an equation for $\lambda^2$, extremization of which leads to eq.(\ref{sturm}) 
\begin{eqnarray}
\lambda^2 &=& \frac{\int_0^1 dz\ \{p(z)[F'(z)]^2 - q(z)[F(z)]^2 \} }
{\int_0^1 dz \ r(z)[F(z)]^2}~.
\label{eq5abc}
\end{eqnarray}
We shall now use the trial function for the estimation of $\lambda^{2}$
\begin{eqnarray}
F= F_{\tilde\alpha} (z) \equiv 1 - \tilde\alpha z^2.
\label{eq50}
\end{eqnarray}
Note that $F$ satisfies the conditions $F(0)=1$ and $F'(0)=0$.\\

\noindent We now proceed to obtain the relation between the critical temperature and the charge density. To do this we start from eq.(\ref{de9}) and use eq.(\ref{lam}). This yields
\begin{eqnarray}
T_c &=& \frac{r_{+(c)}}{4\pi} \left[(d-1) - \frac{4MG_d}{\Gamma(\frac{d-1}{2})} \frac{e^{-\frac{r^2_{+(c)}}{4\theta}}}{(4\theta)^{\frac{d-1}{2}}}\right] \nonumber \\ 
&=& \frac{1}{4\pi} \left[(d-1) - \frac{4MG_d}{\Gamma(\frac{d-1}{2})} \frac{e^{-\frac{(2MG_d)^{\frac{2}{d-1}}}{4\theta}}}{(4\theta)^{\frac{d-1}{2}}}\right]\left(\frac{\rho}{\lambda}\right)^{\frac{1}{d-2}}
\label{de99} 
\end{eqnarray}
where we have used $r^{d-1}_{+(c)}\approx 2MG_d$ (leading order term in eq.(\ref{r1nc}) ) in the exponential term $e^{-\frac{r^2_{+(c)}}{4\theta}}$.
The above result for the critical temperature holds for a $d$-dimensional holographic superconductor and is one of the main results in this paper. For calculating $\lambda^2$, we also have to compute horizon radius $r_{+(c)}$ which can be obtained from $f(r_{+(c)})=0$. It is to be noted that the effect of the BI coupling parameter $b$ in the critical temperature $T_{c}$ enterss through the eigenvalue $\lambda$. For $d=4$, we get 
\begin{eqnarray}
T_c = \frac{3}{4\pi} \left[1 - \frac{MG}{3\sqrt{\pi}\theta^{3/2}}e^{-\frac{(2MG)^{2/3}}{4\theta}}\right] \sqrt{\frac{\rho}{\lambda}} \equiv \xi \sqrt{\rho}
\end{eqnarray}
where $\xi= \frac{3}{4\pi\sqrt{\lambda}}(1 - \frac{MG}{3\sqrt{\pi}\theta^{3/2}} e^{-\frac{(2MG)^{2/3}}{4\theta}})$. This result correctly takes into account the effect of noncommutativity which was not there in \cite{subir}. Note that the effect of $\theta$ is not only entering through $\lambda$, but also the coefficient $\frac{3}{4\pi}$ gets renormalized by a $\theta-$dependent factor. This factor was missing in \cite{subir}.

\noindent In the rest of our analysis, we shall set $d=5$ and $m^2=-3$. The choice for $m^2$ yields $\Delta_{+}=3$ from eq.(\ref{bound1a}). 
Eq.(s)(\ref{de9}, \ref{i1}) therefore becomes
\begin{eqnarray}
\label{nwe}
T_{c}&=&\frac{1}{\pi}\left[ 1- \frac{MG_5}{(4\theta)^2} e^{-\frac{\sqrt{2MG_5}}{4\theta}} \right]\left(\frac{\rho}{\lambda}\right)^{\frac{1}{3}}\\
p(z)&=&z^3 \left(1-z^4 \left\{ 1+ \frac{\sqrt{2MG_5}}{4\theta}e^{-\frac{\sqrt{2MG_5}}{4\theta}} + e^{\frac{\sqrt{2MG_5}}{4\theta}}\right\} \right) \nonumber \\
q(z)&=& -9 z^5 \left\{ 1+ \frac{\sqrt{2MG_5}}{4\theta}e^{-\frac{\sqrt{2MG_5}}{4\theta}} + e^{\frac{\sqrt{2MG_5}}{4\theta}}\right\} \nonumber \\
r(z)&=&\frac{z^3\left\{ (1-z^{2})^2 -b(\lambda^2|_{b=0}) (1-z^{2})(1-z^{8})\right\}}{1-z^4 \left\{ 1+ \frac{\sqrt{2MG_5}}{4\theta}e^{-\frac{\sqrt{2MG_5}}{4\theta}} + e^{\frac{\sqrt{2MG_5}}{4\theta}}\right\}} ~.
\label{de13}
\end{eqnarray}
In the $\theta = 0$ limit (commutative case), $\Xi = 1$ and $T_c = \frac{1}{\pi}\left(\frac{\rho}{\lambda}\right)^{1/3}$. So the values of $T_c$ does not get affected by the mass of black hole. For $\theta=0$, we have $\lambda^2_{\tilde{\alpha}=0.7218}$ = $18.23$ for $b=0$, which can be obtained by following the procedure in \cite{sgdc1}. Using this value of $\lambda^2$, we get $T_c= 0.1962 \rho^{1/3}$ which is in very good agreement with the exact value $T_c=0.1980 \rho^{1/3}$ \cite{hs16}. The effect of Born-Infeld parameter had been studied in arbitrary dimensions for commutative Einstein gravity and Gauss-Bonnet gravity in details \cite{dg1}. In this paper we have included the effect of spacetime noncommutativity along with the effect of Born-Infeld parameter. We have calculated the $T_c$ values for different set of $\theta,~M,~b$ values. We have show that noncommutative parameter is not favourable for condensation.


\noindent In Tables \ref{E1},\ref{E2},\ref{E3}, we present our analytical values obtained by the SL eigenvalue approach for different sets of values of $\theta, ~M$ and $b$.
\begin{table}[htb]
\caption{Analytical results for the critical temperature and the charge density $(\frac{T_c}{\rho^{1/3}})$ with different values of $M$ and $\theta$ for $b=0$(i.e. Maxwell's electrodynamics)}   
\centering                          
\begin{tabular}{|c| c| c| c| c| c| }            
\hline                       
$\theta$& $M=10/G_5$ & $M=50/G_5 $ & $M=100/G_5$  & $M=150/G_5$ & $M=200/G_5$ \\
\hline
$\theta= 0.3$ & 0.1638 & 0.1946 & 0.1961 & 0.1964 & 0.1962 \\ 
$\theta= 0.5$ & 0.1468 & 0.1798 & 0.1921 & 0.1949 & 0.1958   \\
$\theta= 0.7$ & 0.1540 & 0.1615 & 0.1803 & 0.1885 & 0.1923  \\
$\theta= 0.9$ & 0.1686 & 0.1505 & 0.1667 & 0.1778 & 0.1845 \\
\hline                 
\end{tabular}
\label{E1}  
\end{table}
\begin{table}[htb]
\caption{Analytical results for the critical temperature and the charge density $(\frac{T_c}{\rho^{1/3}})$ with different values of $\theta$ and $b$ for fixed value $M= 100/G_5 $}   
\centering                          
\begin{tabular}{|c| c| c| c| c| }            
\hline                       
$b $ & $\theta= 0.3$ & $\theta= 0.5$ & $\theta= 0.7$  & $\theta= 0.9$ \\
\hline
$b= 0.01$ & 0.1850 & 0.1811 & 0.1700 & 0.1574 \\ 
$b= 0.02$ & 0.1694 & 0.1658 & 0.1557 & 0.1445 \\
\hline                 
\end{tabular}
\label{E2}  
\end{table}
\begin{table}[htb]
\caption{Analytical results for the critical temperature and the charge density $(\frac{T_c}{\rho^{1/3}})$ with different values of $M$ and $b$ for fixed value $\theta= 0.5 $}   
\centering                          
\begin{tabular}{|c| c| c| c| c| c| }            
\hline                       
$b $ & $M=10/G_5$ & $M=50/G_5 $ & $M=100/G_5$  & $M=150/G_5$ & $M=200/G_5$ \\
\hline
$b= 0.01$ & 0.1390 & 0.1696 & 0.1811 & 0.1838 & 0.1846 \\ 
$b= 0.02$ & 0.1284 & 0.1556 & 0.1658 & 0.1683 & 0.1690 \\
\hline                 
\end{tabular}
\label{E3}  
\end{table}



\section{Condensation operator and critical exponent}
To investigate the relation between the condensation operator and the critical temperature, we look at the field equation (\ref{e1aa}) for $\phi (z)$ near to the critical temperature $(T_{c})$  
\begin{eqnarray}
\phi^{\prime \prime}(z) - \frac{d-4}{z} \phi^{\prime}(z) + \frac{d-2}{r^2_{+}} b \phi^{\prime}(z)^{3} z^3 = \frac{\langle J\rangle^{2}}{r^{2}_{+}} \mathcal{B}(z)\phi (z)
\label{cx1} 
\end{eqnarray}
where $\mathcal{B}(z)= \frac{2 z^{2\Delta_{+} -4}}{r_{+}^{2\Delta_{+} -4}}\frac{F^{2}(z)}{f(z)}\left( 1- \frac{b z^4}{r^2_{+}}\phi^{\prime}(z)^2 \right)^{\frac{3}{2}}.$

\noindent We may now expand $\phi(z)$ in the small parameter $\frac{\langle J\rangle^2}{r^{2}_{+}}$ as 
\begin{eqnarray}
\frac{\phi(z)}{r_{+}} = \lambda \left\{ (1-z^{d-3}) - \frac{b(\lambda^2|_{b=0}) (d-3)^3}{2(3d-7)} (1-z^{3d-7})\right\} + \frac{\langle J\rangle^2}{r^{2}_{+}} \zeta (z)
\label{cx2} 
\end{eqnarray}
with $\zeta (1)= 0 =\zeta^{\prime}(1).$\\
Using eq.(\ref{cx2}) and comparing the coefficient of $\frac{\langle J\rangle^2}{r^{2}_{+}}$ on both sides of eq.(\ref{cx1}) (keeping terms upto $\mathcal{O}(b)$), we get the equation for the correction $\zeta(z)$ near to $T_c$
\begin{eqnarray}
\zeta^{\prime \prime}(z) -\left\{ \frac{d-4}{z} + 3b(\lambda^2|_{b=0}) (d-2)(d-3)^2 z^{2d-5} \right\} \zeta^{\prime}(z) = \lambda \frac{2 z^{2\Delta_{+} -4}}{r_{+}^{2\Delta_{+} -4}}\frac{F^{2}(z)}{f(z)}\mathcal{A}_{1} (z)
\label{cx3}
\end{eqnarray}
where $\mathcal{A}_{1} (z) = 1-z^{d-3} -\frac{3b(\lambda^2|_{b=0}) (d-3)^2}{2}\left\{(1-z^{d-3})z^{2d-4} +\frac{d-3}{3(3d-7)}(1-z^{3d-7}) \right\} $. \\
To solve this equation, we need to multiply this equation by its integrating factor which turns out to be $z^{-(d-4)} e^{\frac{3(d-2)(d-3)^2 b(\lambda^2|_{b=0})}{2d-4} z^{2d-4}} $. This leads to (after using eq.(\ref{metr33}))
\begin{eqnarray}
\frac{d}{dz}\left( z^{-(d-4)} e^{\frac{3(d-2)(d-3)^2 b(\lambda^2|_{b=0})}{2d-4} z^{2d-4}} \zeta^{\prime}(z) \right)= \lambda \frac{2 z^{2\Delta_{+} -2}}{r_{+}^{2\Delta_{+} -2}}\frac{z^{-(d-4)} F^{2}(z)}{(1-\Xi. z^{d-1})}  e^{\frac{3(d-2)(d-3)^2 b(\lambda^2|_{b=0}) z^{2d-4}}{2d-4} } \mathcal{A}_{1} (z). \nonumber \\
\label{cx4}
\end{eqnarray}
Using boundary condition of $\zeta(z)$, we integrate eq.(\ref{cx4}) between the limits $z=0$ and $z=1$. This yields
\begin{eqnarray}
\frac{\zeta^{\prime}(z)}{z^{d-4}}\mid_{z\rightarrow 0} = -\frac{\lambda}{r^{2\Delta_{+} -2}_{+}} \mathcal{A}_{2}
\label{cx5}
\end{eqnarray}
where $\mathcal{A}_{2} = \int^{1}_{0} dz \frac{2 z^{2\Delta_{+}-2}z^{-(d-4)} F^{2}(z)}{(1-\Xi. z^{d-1})}  e^{\frac{3(d-2)(d-3)^2 b(\lambda^2|_{b=0})}{2d-4} z^{2d-4}} \mathcal{A}_{1} (z) $.\\

\noindent Now the asymptotic behaviour of $\phi(z)$ is given by eq.(\ref{bound1}). But from eq.(\ref{cx2}), we get the asymptotic behaviour (near $z=0$) of $\phi(z)$. This leads to the following equation: 
\begin{eqnarray}
\mu -\frac{\rho}{r^{d-3}_{+}}z^{d-3} &=& \lambda r_{+} \left\{ (1-z^{d-3}) - \frac{b(\lambda^2|_{b=0}) (d-3)^3}{2(3d-7)} (1-z^{3d-7})\right\} \nonumber \\
&+& \frac{\langle J\rangle^2}{r_{+}} \left\{\zeta(0)+z\zeta^{\prime}(0)+......+\frac{\zeta^{d-3}(0)}{(d-3)!} z^{d-3}+....\right\}
\label{cx7}
\end{eqnarray}
Comparing the coefficient of $z^{d-3}$ on both sides of the above equation, we obtain
\begin{eqnarray}
-\frac{\rho}{r^{d-3}_{+}} = -\lambda r_{+} + \frac{\langle J\rangle^2}{r_{+}}.\frac{\zeta^{d-3}(0)}{(d-3)!}
\label{cx8}
\end{eqnarray}
together with $ \zeta^{\prime}(0)= \zeta^{\prime \prime}(0)=.........= \zeta^{d-4}(0)= 0$.\\

\noindent Now we note that $\zeta^{\prime}(z)$ and $(d-3)^{th}$ derivative of $\zeta(z)$ are related by
\begin{eqnarray}
\frac{\zeta^{d-3}(z=0)}{(d-4)!} = \frac{\zeta^{\prime}(z)}{z^{d-4}}|_{z\rightarrow 0}
\label{cx6}
\end{eqnarray}
when $\zeta^{\prime}(0)= \zeta^{\prime \prime}(0)=.........= \zeta^{d-4}(0)= 0$ and $ d \geq 4 $.\\
Using this, the eq.(\ref{cx8}) gives the relation between the charge density $(\rho)$ and the condensation operator $(\langle J\rangle)$ 
\begin{eqnarray}
\frac{\rho}{r^{d-2}_{+}} = \lambda\left[1 + \frac{\langle J\rangle^2}{r^{2\Delta_{+}}_{+}}.\frac{\mathcal{A}_{2}}{(d-3)} \right]
\label{cx9}
\end{eqnarray}  
Using eq.(s)(\ref{de9},\ref{lam}) and simplifying eq.(\ref{cx9}), we get
\begin{eqnarray}
\langle J\rangle^2 = \frac{(d-3)(4\pi T_{c})^{2\Delta_{+}}}{\mathcal{A}_{2}\left[(d-1)- \frac{4MG_d}{\Gamma(\frac{d-1}{2})}.\frac{e^{-\frac{(2MG_d)^{2/(d-1)}}{4\theta}}}{(4\theta)^{\frac{d-1}{2}}}\right]^{2\Delta_{+}}}.\left(\frac{T_{c}}{T}\right)^{d-2} \left[1- \left(\frac{T}{T_{c}}\right)^{d-2} \right]
\label{cx10}
\end{eqnarray}
Note that the temperature is away from (but close to) the critical temperature (i.e. $T \approx T_{c}$) and therefore we can write
\begin{eqnarray}
\left(\frac{T_{c}}{T}\right)^{d-2} \left[1- \left(\frac{T}{T_{c}}\right)^{d-2} \right] &=& \left(\frac{T_{c}}{T}\right)^{d-2}\left[1- \left(\frac{T}{T_{c}}\right)\right]\left[1+\frac{T}{T_{c}} + \left(\frac{T_{c}}{T}\right)^2 +.....+\left(\frac{T_{c}}{T}\right)^{d-3} \right] \nonumber \\
&\approx & (d-2)\left[1- \left(\frac{T}{T_{c}}\right)\right].
\label{cx11}
\end{eqnarray} 
Substituting the above relation in eq.(\ref{cx10}) leads to expression between the condensation operator and the critical temperature in $d-$dimension
\begin{eqnarray}
\langle J\rangle = \beta T^{\Delta_{+}}_{c} \sqrt{1-\frac{T}{T_{c}}}
\label{cx12}
\end{eqnarray}
where $\beta = \sqrt{\frac{(d-3)(d-2)}{\mathcal{A}_{2}}}.(4\pi)^{\Delta_{+}}.\left[(d-1)- \frac{4MG_d}{\Gamma(\frac{d-1}{2})}.\frac{e^{-\frac{(2MG_d)^{2/(d-1)}}{4\theta}}}{(4\theta)^{\frac{d-1}{2}}}\right]^{-\Delta_{+}} $.\\

\noindent As a byproduct of our analysis, we find that the critical exponent is $1/2$ which is not affected by the noncommutativity of spacetime. \\
We now perform some explicitly computation with $d=5, ~m^2=-3$. The choice for $m^2$ yields $\Delta_{+}=3$. Eq.(\ref{cx12}) then becomes 
\begin{eqnarray}
\langle J\rangle = \beta T^{3}_{c} \sqrt{1-\frac{T}{T_{c}}}
\label{cx13}
\end{eqnarray}
In $d=5$-dimension,
\begin{eqnarray}
\mathcal{A}_{1} (z) &=& (1-z^2)\left[1-\frac{b(\lambda^2|_{b=0})}{2} (1+z^2 +z^4 +13z^6) \right] \nonumber \\
\mathcal{A}_{2} &=& \int^{1}_{0} dz \frac{2 z^{3} F^{2}(z)}{(1-\Xi. z^{4})}  e^{6 b(\lambda^2|_{b=0}) z^{6}} \mathcal{A}_{1}(z) \nonumber\\
&\approx& \int^{1}_{0} dz \frac{2 z^{3} F^{2}(z) (1-z^2)}{(1-\Xi. z^4)} \left\{1- \frac{b(\lambda^2|_{b=0})}{2} (1+z^2 +z^4 +z^6) \right\} \nonumber \\
\beta &=& \sqrt{\frac{6}{\mathcal{A}_{2}}} \left[\frac{\pi}{1-\frac{MG_5}{(4\theta)^2}.e^{-\frac{\sqrt{2MG_5}}{4\theta} }} \right]^{3} =\beta_1 \pi^3
\label{cx14}
\end{eqnarray}
For $\theta= 0$, we recover the commutative result $\beta_{1}= 7.705$ which is very good agreement with exact value $\beta_{1}=7.706$ \cite{hs24}.

\noindent In Tables \ref{E4}, \ref{E5}, \ref{E6}, we present our analytical values of $\beta_1$ obtained by the SL eigenvalue approach for different sets of values of $b, ~M $ and $\theta$.
\begin{table}[ht]
\caption{Analytical results for the condensation operator values $\beta_1$ with different values of $M$ and $\theta$ for $b=0$(i.e. Maxwell's electrodynamics)}   
\centering                          
\begin{tabular}{|c| c| c| c| c| c| }            
\hline                       
$\theta$& $M=10/G_5$ & $M=50/G_5 $ & $M=100/G_5$  & $M=150/G_5$ & $M=200/G_5$ \\
\hline
$\theta= 0.3$ & 14.265 & 7.914 & 7.728 & 7.726 & 7.795 \\ 
$\theta= 0.5$ & 23.645 & 10.265 & 8.253 & 7.862 & 7.763   \\
$\theta= 0.7$ & 17.784 & 15.024 & 10.172 & 8.762 & 8.226  \\
$\theta= 0.9$ & 13.608 & 20.057 & 13.371 & 10.663 & 9.407 \\
\hline                 
\end{tabular}
\label{E4}  
\end{table}
\begin{table}[htb]
\caption{Analytical results for the condensation operator values $\beta_1$ with different values of $\theta$ and $b$ for fixed value $M= 100/G_5 $}   
\centering                          
\begin{tabular}{|c| c| c| c| c| }            
\hline                       
$b $ & $\theta= 0.3$ & $\theta= 0.5$ & $\theta= 0.7$  & $\theta= 0.9$ \\
\hline
$b= 0.01$ & 8.755 & 9.340 & 11.497 & 14.984 \\ 
$b= 0.02$ & 10.380 & 11.104 & 13.562 & 17.501 \\
\hline                 
\end{tabular}
\label{E5}  
\end{table}
\begin{table}[t!]
\caption{Analytical results condensation operator values $\beta_1$ with different values of $M$ and $b$ for fixed value $\theta= 0.5 $}   
\centering                          
\begin{tabular}{|c| c| c| c| c| c| }            
\hline                       
$b $ & $M=10/G_5$ & $M=50/G_5 $ & $M=100/G_5$  & $M=150/G_5$ & $M=200/G_5$ \\
\hline
$b= 0.01$ & 25.721 & 11.604 & 9.340 & 8.917 & 8.801 \\ 
$b= 0.02$ & 28.715 & 13.679 & 11.071 & 10.572 & 10.440 \\
\hline                 
\end{tabular}
\label{E6}  
\end{table}


\section{Conclusions}
In this paper, we have investigated the role of noncommutativity of spacetime in holographic superconductors in the framework of BI electrodynamics. In our analysis, we have obtained the relation between the critical temperature and the charge density in $d-$dimensions. We observe that critical temperature not only depends on the charge density but also on the noncommutative parameter $\theta$ and the mass of the black hole. We have presented the expressions of $T_c$ for $d=4, ~5$. We have analytically calculated critical temperature $T_c$ and the condensation operator $\langle J \rangle$ for $d=5$. Our analytic results show  that the condensation gets harder to form in the presence of Born-Infeld parameter and the noncommutativity of spacetime. It is also observed that the formation of the condensate is favoured for large black hole mass. As a future work, we would like to study the set up away from the probe limit for Gauss-Bonnet gravity.

\section*{Acknowledgments} DG would like to thank DST-INSPIRE, Govt. of India for financial support. DG would also like to thank Prof. Biswajit Chakraborty of S.N.Bose Centre for constant encouragement.
S.G. acknowledges the support by DST SERB under Start Up Research Grant (Young Scientist), File No.YSS/2014/000180.


\end{document}